# Evaluation of a front-end ASIC for the readout of PMTs in large dynamic range

Wu Wei-Hao(邬维浩)[1] Zhao Lei(赵雷)[1;1)] Liang Yu(梁宇)[1] Yu Li(于莉)[1]

Liu Jian-Feng(刘建峰)[1] Liu Shu-Bin(刘树彬)[1] An Qi(安琪)[1]

[1]State Key Laboratory of Particle Detection and Electronics, USTC, Hefei, 230026

**Abstract:** The Large High Altitude Air Shower Observatory (LHAASO) project has been proposed for the survey and study of cosmic rays. In the LHAASO project, the Water Cherenkov Detector Array (WCDA) is one of major detectors for searching gamma ray sources. A Charge-to-Time Convertor (QTC) ASIC (Application Specification Integrated Circuit) fabricated in Global Foundry 0.35 μm CMOS technology, has been developed for readout of Photomultiplier Tubes (PMTs) in the WCDA. This paper focuses on the evaluation of this front-end readout ASIC performance. Test results indicate that the time resolution is better than 400 ps and the charge resolution is better than 1% with large input signals and remains better than 15% @ 1 Photo Electron (P.E.), both beyond the application requirement. Moreover, this ASIC has a weak ambient temperature dependence, low input rate dependence and high channel-to-channel isolation.

## 1 Introduction

The Large High Altitude Air Shower Observatory (LHAASO) project oriented to the study and exploration of particle astrophysics has been proposed, consisting of various kinds of efficient detectors [1-2]. One of the major components is the Water Cherenkov Detector Array (WCDA) aiming to survey gamma ray origins with its high sensitivity to gamma ray showers above a few hundred GeV[3-4]. The WCDA is composed of four 150m×150m water ponds, each with 900 Photomultiplier Tubes (PMTs) located at the bottom to collect Cherenkov light produced in the water by secondary particles of an air shower. A total of 3600 electronics readout channels are required, as shown in Fig. 1. Readout electronics of every 9 adjacent PMTs are combined into one Front-End Electronics module (FEE), which processes and digitizes the PMT output signals before sending them to the Data Acquisition (DAQ) system [5-8].

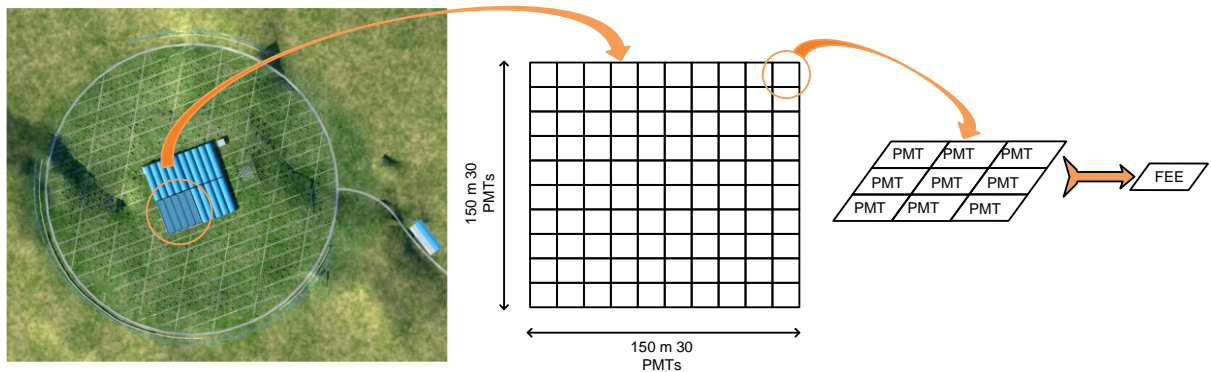

Fig. 1. WCDA in LHAASO project and its readout scheme.

---

1) zlei@ustc.edu.cn

To target the research purposes, both the amount of the Cherenkov light and their arrival time on the PMTs need to be efficiently measured in a large dynamic range. The readout electronics for the WCDA are required to record the leading edge and charge of the PMT signals in a range from single P.E. (Photo Electron) to 4000 P.E. The resolution of time measurement should be better than 500 ps in the full dynamic rang. And the charge resolution is required to be better than 3% with large input signals, and better than 30% @ 1 P.E. The detailed requirements of the readout electronics for WCDA are listed in Table 1.

Table 1.   Requirement of the Readout Electronics

| Item | Requirement |
| --- | --- |
| Channel Number | 3600 |
| Dynamic Range | 1PE~4000PE |
| Time Resolution | 500 ps |
| Charge Resolution | 3% @4000 PE; 30% @ 1 PE |
| Average Rate | 50 kHz |

There are some high performance readout ASICs that have been designed for PMTs in other experiments [9-12]. However, their performance does not meet the requirement of WCDA in the LHASSO. In order to fulfill the large dynamic range of charge measurement, we employ two readout channels to record the signals from the anode and 10th dynode of each PMT respectively, as shown in Fig. 2. The anode channel measures PMT signals in a low range of 1~100 P.E., while the 10th dynode channel covers a range of 40~4000 P.E. In this case, a full dynamic range of 4000 is achieved with a sufficient overlap.

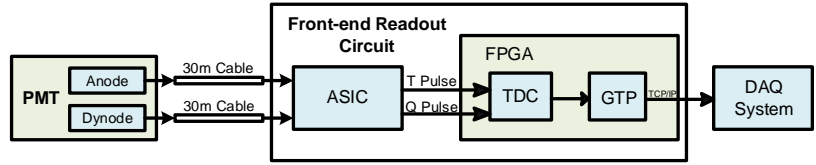

Fig. 2. Architecture of readout electronics for PMTs in WCDA.

Besides the large dynamic range, the main challenge for the readout electronics is the precise time and charge measurement at 1 P.E. level. The waveform of a single P.E. signal from the anode of PMT (R5912) is shown as the red one in Fig. 3, with a 4.6 ns rising time, 16 ns falling time, and 3 mV peak amplitude with 50 Ω terminate resistor. Moreover, the PMT signals are further attenuated during transmission from the PMT with 30-meter coaxial cables, as shown in the blue one of Fig. 3.

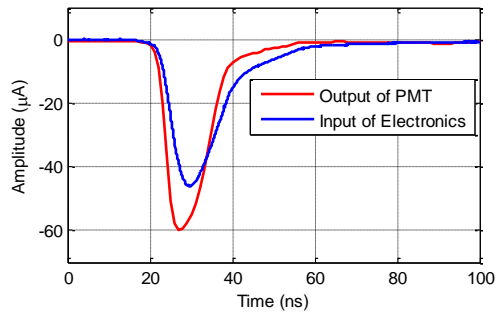

Fig. 3. The waveform of single P.E. signal from PMT.

## 2 ASIC Structure

In order to simplify the front electronics, we have designed a front-end Charge-to-Time Convertor (QTC) ASIC, integrating all front end analog circuits in it. It is fabricated in Global Foundry 0.35 μm CMOS technology and packaged in a 100-pin plastic LQFP. This QTC ASIC processes the signals from anode and 10th dynode of PMTs: discriminating and converting the charge information into the output pulse width. Comparing to the traditional method based on pre-amplification, shaping & A/D conversion, only a Field Programmable Gate Array (FPGA) based Time-to-Digital Converter (TDC) and some peripheral components are required in the FEE, decreasing the complexity of the electronics system dramatically.

As the PMT signals are sent to the readout electronics via coaxial cables of 30 meter, precise impedance match should be achieved to suppress signal reflection. Moreover, protection circuits should also be considered. The external input circuits for the QTC ASIC are shown in Fig. 4. The one for the anode channel consists of a termination resistor $R_{t1}$, a voltage limiting resistor $R_{l1}$, a diode and an AC coupling capacitor. In Fig. 4, $C_{p1}$ refers to the equivalent parasitic capacitor of the QTC ASIC, the package, and the transmission lines of the test board. As for the dynode readout channel, a 20 dB Pi-pad attenuator is applied in front of the termination resistor.

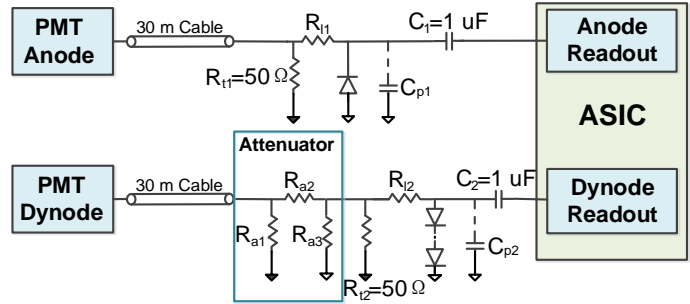

Fig. 4. External circuit for the QTC ASIC.

The diagram of the anode readout channel inside the QTC ASIC is shown in Fig. 5. The input signals are amplified by a low-noise amplifier before being sent into two separate measurement parts: time measurement and charge measurement. In the time measurement part, the voltage signals are amplified by another low-noise amplifier, processed by a Voltage-to-Current (VI) convertor and discriminated by a current discriminator. The output of the current discriminator controls the current integration process of the charge measurement part, as well as generates "T Output" in Fig. 5 for time measurement. Once the integration process is finished, the charge on the capacitor is discharged by a constant current source. In this case, the duration of the discharge process has a linear relationship with the input charge information. Applying a voltage discriminator and some logic circuits, a pulse signal ("Q Output") could be generated, with its width corresponding to the charge information of input signals. As for the dynode readout channel, the circuit structure is quite similar. Considering different polarity and large amplitude of the input signals, an invert buffer is implemented in the dynode readout channel.

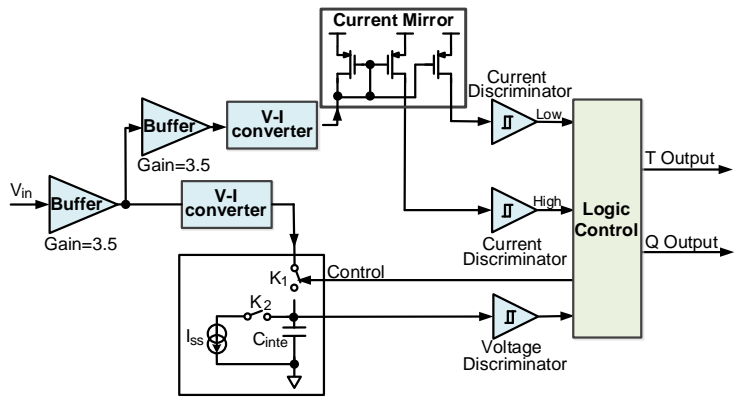

Fig. 5. Diagram of the anode readout channel.

The simulated waveforms of critical signals of the anode channel are shown in Fig. 6. From the up to down, these four signals are input signal, output of current discriminator, charge output pulse and voltage on the capacitor.

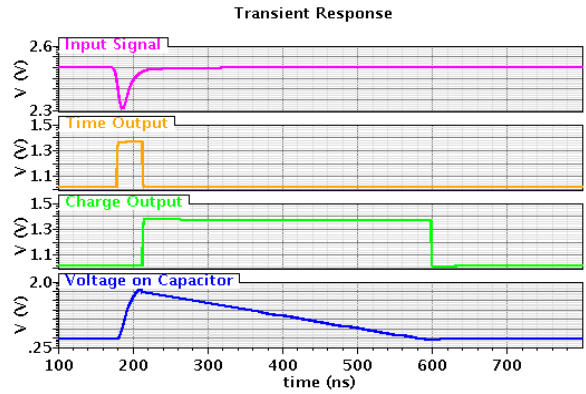

Fig. 6. Waveforms of signals in the anode channel.

## 3 Test results

### 3.1 Test Platform

In order to evaluate the performance of the QTC ASIC, a test board was designed. The test platform is shown in Fig. 7; it consists of a power supply, a signal source (Agilent Technologies 81160A), a wide bandwidth attenuator (Wavetek Step Attenuator Model 5080.1), a coaxial cable of 30 meter, an oscilloscope (Lecroy 104 MXi) and the test board with the QTC ASIC mounted. The signal source generates detector-like signals with a repetition rate of 50 kHz. The attenuator is used to adjust the amplitude of input signals while maintaining a good signal to noise ratio.

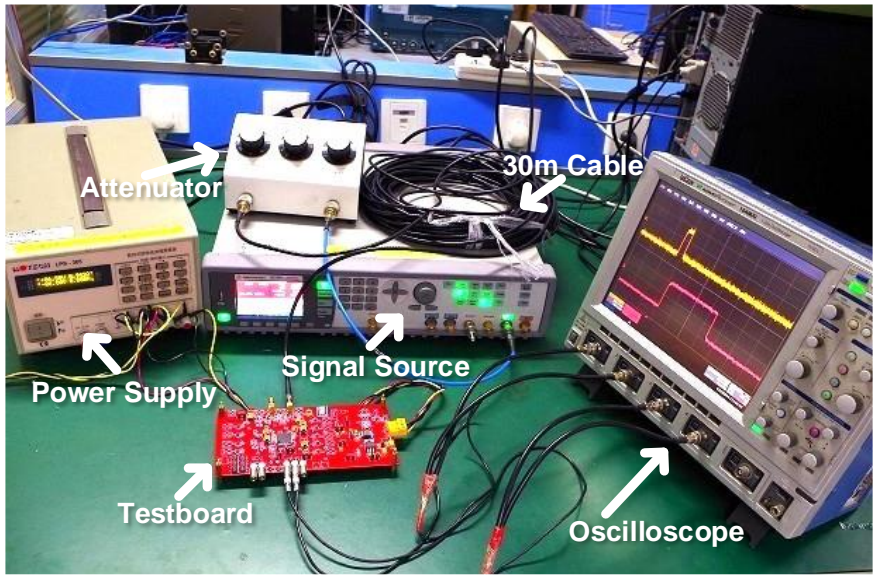

Fig 7. Test platform of the QTC ASIC.

**3.2 Functional Test**

The functional test was conducted by observing the transient waveform of the critical points in the QTC ASIC, as shown in Fig. 8. The recorded waveforms correspond to those in the simulation results in Fig. 6, which concord well. Then, a series of tests were conducted to evaluate the performance of the ASIC: time and charge resolution, dynamic ranges, crosstalk between different channels, ambient temperature dependence and input rate dependence. The tests results are presented as follows.

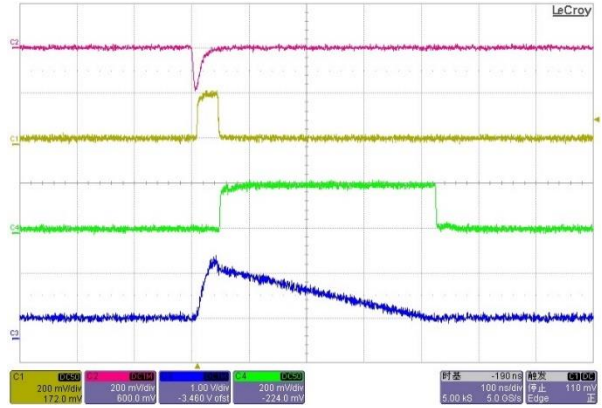

Fig. 8. Waveforms of the functional test (red: input signal, yellow: output of current discriminator, green: charge output, blue: voltage on the integration capacitor).

**3.3 Timing Measurement Performance**

The timing performance is the critical requirement for readout electronics of PMTs. To test the time resolution, we measured the RMS of the time interval between the time output signal from the QTC ASIC and a reference signal from the signal source. The test results of time resolution are shown in Fig. 9 (a). It can be observed that the time resolution is about 300 ps with single P.E. signal input, and it is better than 100 ps with input signals larger than 10 P.E. The time walk is shown in Fig. 9 (b). Large signals from PMTs reach the threshold earlier than small signals, due to the front edge discrimination applied. The time walk is smaller than 10 ns in the full measurement range, which can be further calibrated with charge measurement results.

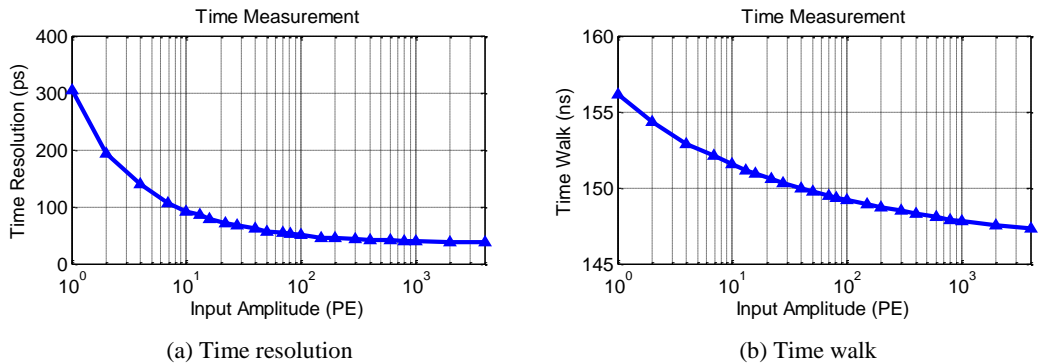

(a) Time resolution  (b) Time walk

Fig. 9. Timing performance test results.

During the test, we found that the parasitic capacitance of external input circuits in the test board has great impact on the timing performance of the QTC ASIC. As for external input circuits of the anode readout channel shown in Fig. 4, the voltage limiting resistor $R_{l1}$ not only contributes serial noise but also decreases the bandwidth of the input signals together with the equivalent parasitic capacitor $C_{p1}$. To assess the influence of the resistor $R_{l1}$ and parasitic capacitor $C_{p1}$, we conducted simulation, and compared with and the test results.

We recorded the signals across the resistor $R_{l1}$, as shown in Fig. 10. We can find that the input signal is low-pass filtered, with amplitude attenuation of around 0.66. This is due to the one order low pass filtered formed by $R_{l1}$ and parasitic $C_{p1}$ shown in Fig. 4. (In this test, the strobe of the oscilloscope contributes parasitic capacitance of 8 pF.)

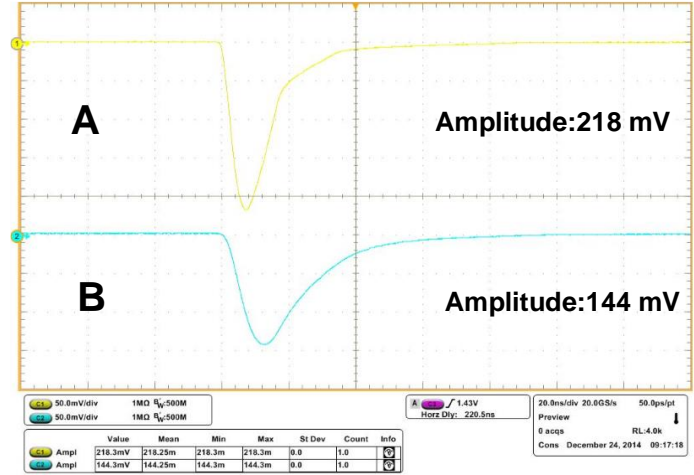

Fig. 10. Waveforms of signals on the voltage limiting resistor (A: signal of input end of the resistor; B: signal of output end of the resistor).

We also simulate response of the RC filter with the PMT signal input. The relation between the amplitude attenuation and the capacitor value is shown in Fig.11. Deducting the oscilloscope's strobe 8 pF parasitic capacitance, we can know that the value of the parasitic capacitor $C_{p1}$ is about 11.4 pF, according to both the test results and simulation results. The parasitic capacitor $C_{p1}$ consists of the input equivalent capacitor of the QTC ASIC ($C_{p1\_input}$; 3 pF in the simulation results) and the parasitic capacitor ($C_{p1\_test}$) in the test board (stripe lines and pads between the voltage limiting resistor $R_{l1}$ and the input of the QTC ASIC). In this case, we can find that the test board parasitic capacitor $C_{p1\_test}$ is about 8.4 pF. These results indicate the distinct influence of the parasitic capacitor. This gives direction to further improve the timing performance in next version: choosing the package type with lower input parasitic capacitance like QFN, reducing the length of stripe lines in PCB and so on.

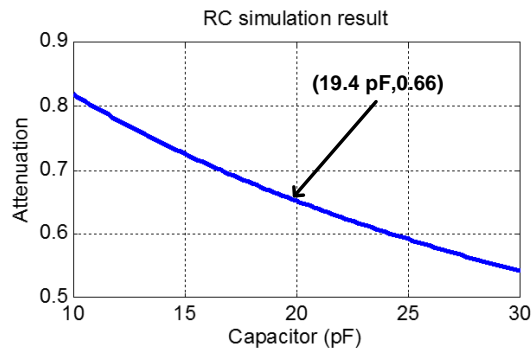

Fig. 11 RC simulation results.

### 3.4 Charge Measurement Performance

The charge information of the PMT signals (related to the amount of Cherenkov light) is converted to the width of a pulse signal in the QTC ASIC. We tested the anode channel and dynode channel, respectively. Their charge pulse width are shown in Fig. 12 (a) as a function of input signals. Each channel has a dynamic range of about 100. As shown in Fig. 12 (b), the charge resolution is better 15% with single P.E. signal input, and it is better than 1% for large input signals.

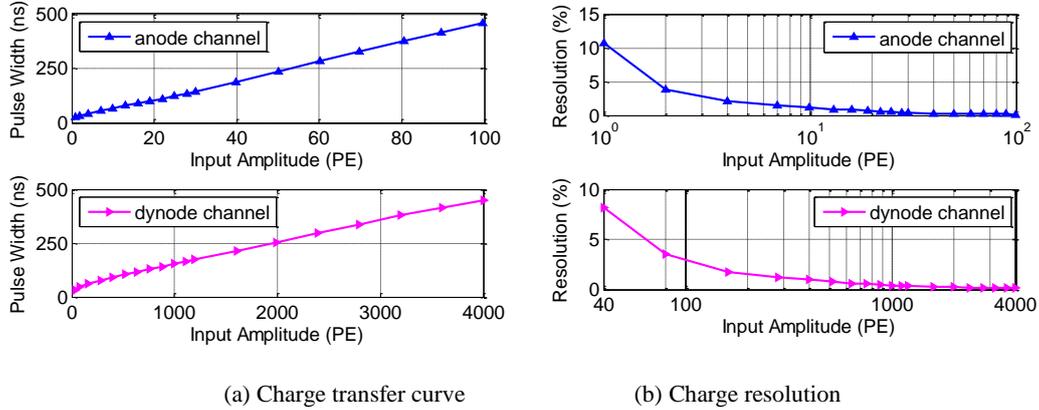

(a) Charge transfer curve    (b) Charge resolution

Fig.12 Charge Measurement test results.

### 3.5 Crosstalk

The crosstalk was also tested by counting error discrimination outputs when large signals are imported to the neighboring channels. The test results showed no error hits has been observed. Moreover, no obvious performance deterioration was observed with or without large signals imported into neighboring channels. It is indicated that the QTC ASIC has a good channel-to-channel isolation.

### 3.6 Ambient Temperature Dependence

We test the ambient temperature dependence of this QTC ASIC in a temperature range from 0 ℃ to 50 ℃ which covers the temperature variation in the real application. The temperature dependence of time resolution is shown in Fig. 13. The time resolution is better than 400 ps in the full temperature range. The timing performance with single P.E. signal input decreases when temperature increases; this is due to that the electronic noise has a proportional relationship with the ambient temperature. The temperature dependence of charge transfer curve is shown in Fig. 14 (a) which can be further calibrated in the system level. As shown in Fig. 14 (b), the charge resolution of single P.E. signal remains better than 15%, which exceed the design requirement of 30%.

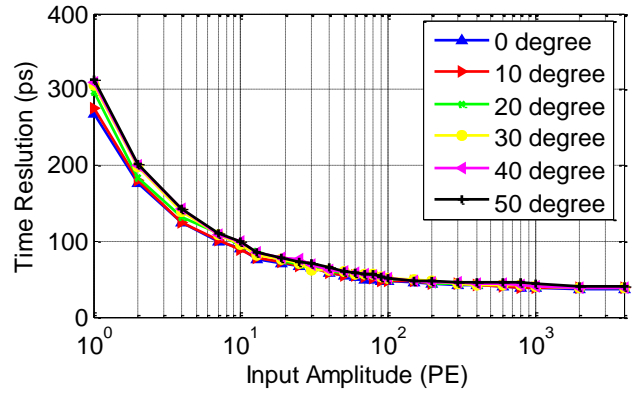

Fig.13 Temperature dependence of time resolution.

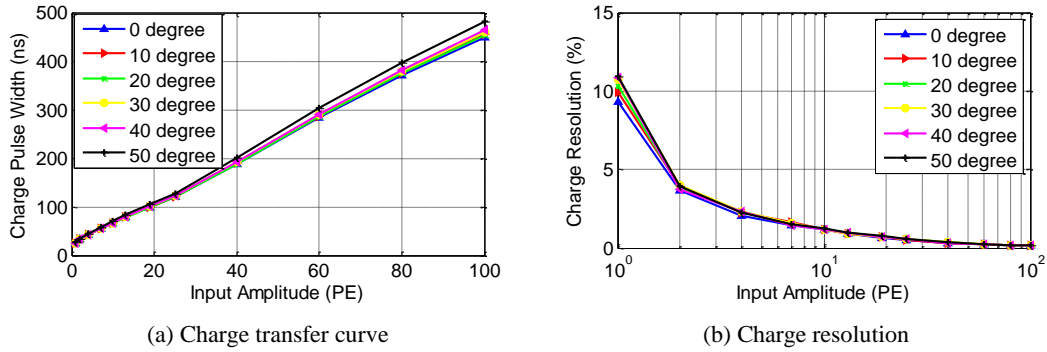

(a) Charge transfer curve    (b) Charge resolution

Fig 14. Temperature dependence of charge measurement.

## 3.7 Input Rate Dependence

The input rate dependence of the QTC ASIC relates to the baseline restoration. It is critical for processing of high update rate signals, especially for charge measurement. Generally, the QTC ASIC have different input rate dependence for input signals with different amplitudes. Fig. 15 shows the QTC ASIC's charge output pulse width vs. the input rate with 1 P.E., 10 P.E. and 100 P.E. input signals. We can see that this ASIC has an input rate dependence of smaller than 5% in the range from 1 kHz to 500 kHz. And it achieves 1% in the range from 1 kHz to 100 kHz which covers the average input rate of 50 kHz.

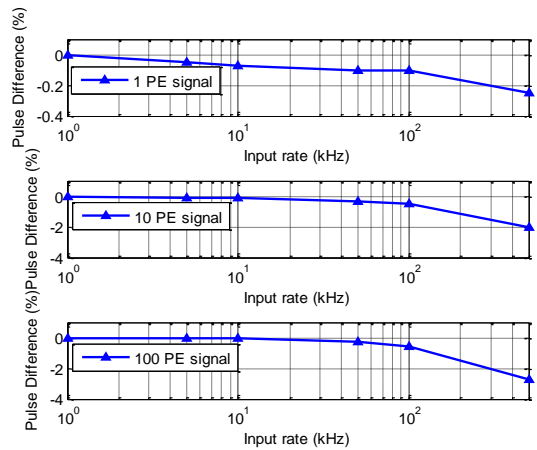

Fig. 15 Input rate dependence test results.

# 4 Conclusion

A front-end ASIC has been developed for the readout of PMTs in the WCDA of LHAASO. It features precise time and charge measurement over a large dynamic range. The kernel performance of the ASIC has been evaluated through a series of tests in the laboratory. Test results indicate that the time resolution is better than 400 ps and the charge resolution is better than 1% with large input signals and remains better than 15% @ 1 P.E., both beyond the application requirement. Moreover, this ASIC has a weak ambient temperature dependence, low input rate dependence and high channel-to-channel isolation.

**Acknowledgement**

The authors would like to thank Allan K. Lan of the MD Anderson Cancer Center in University of Texas for his help regarding this work over the years.

________________________